\documentclass[twocolumn,prb,aps,superscriptaddress,showpacs]{revtex4}
\usepackage{graphicx}
\usepackage{epsfig}
\usepackage{amsmath,amssymb}
\usepackage{natbib}
\begin{document}
\title{One-band Hubbard model with hopping
asymmetry and the effect theory at finite $U$: Phase diagram and
metal-insulator transition}
\author{Yuchuan Wen}
\author{Yue Yu}
\affiliation{Institute of Theoretical Physics, Chinese Academy of
Sciences, P.O. Box 2735, Beijing 100080, China}

\date{\today}

\begin{abstract}
We study the one-band Hubbard model at half filling with hopping
asymmetry and its effective model at finite but large $U$ up to
the second order of t/U. Two variational wave functions, the
resonating valence bond (RVB) wave function and
anti-ferromagnetic-RVB(AF-RVB) coexisted wave function, are
studied by variational Monte Carlo method on L$\times$L square
lattices up to L=12. Based on these two wave functions, the phase
diagrams for both models are presented. For the Hubbard model, we
find that there is a metal-insulator transition when the hopping
parameter $t_{mix}$ which changes the local double occupant
vanishes while only a metal-insulator crossover is explored for
any finite $t_{mix}$. For the effective model in which the
perturbation expansion is up to the second order of $t_{mix}/U$, a
clear metal-insulator transition can be identified for both
variational wave functions and the phase diagram can be drawn
accordingly. In both models, we find that the systems are
dominated by AF-RVB wave function when U is large while RVB wave
function is favored when U is small.
\end{abstract}


\maketitle

\section{Introduction}

The one-band Hubbard model is a generic model for interacting
electrons in the narrow-band and strongly correlated systems
\cite{Hubb}. Especially, since the high temperature
superconductivity was discovered in the cuprates, the Hubbard
model on two-dimensional lattices as well as its strong coupling
limit model \cite{andzr}, the t-J model, have been extensively
studied  in order to understand the various anomalous properties
of the cuprate superconductor. The up-date investigations,
however, can not supply a definitive evidence to show the stable
$d$-wave superconducting ground state in these strongly correlated
models \cite{lau}.

It was known that since the Hubbard model for two-dimensions is
not exactly solvable, using the Hubbard model to study the
exchange correlation of the system is difficult. On the other
hand, the t-J model is weaker in studying the long range charge
correlation because the on-site Coulomb repulsion becomes trivial
due to the non-doubly occupied projection. A better
phenomenological model to include stronger correlations is the
t-J-U model \cite{hua}. The existence of both J- and U- terms is
very important in a possible new mechanism of superconductivity,
gossamer superconductivity, proposed by Laughlin recently
\cite{lau,zhang,lau1}. Both J and U appearing in the model has
been argued to be the result of the correlations like charge
transfer processes in the three band Hubbard model \cite{daul}.
Dealing with the three-band Hubbard model, however, is very
complicated and thus, the precise analytical deduction from the
three band to single band Hubbard models lacks. Moreover, an exact
reduction from the three band model to the single band with both J
and U terms are more difficult.

In this paper, we would like to deal with the one-band Hubbard
model with a hopping asymmetry at finite U. In this case, since
the on-site Coulomb repulsion is not infinite, there may be a
fraction of the lattice sites doubly occupied by electrons. Thus,
the U-term is non-trivial but can be exactly treated. The
difficulty is to deal with the kinetic term. We will present a
variational Monte Carlo calculation for the Hubbard model in a
two-dimensional square lattice. We examine two types of
variational wave functions, the resonating valence bond (RVB) wave
function and anti-ferromagnetic-RVB (AF-RVB) coexisted wave
function. There are many variational wave functions, including AF,
RVB and AF-RVB, due to different kind of approximation. The early
studies shows that the results of AF and RVB were contradictory
but the AF-RVB had lower energy{\cite{Giamarchi,og}}. So it is
quite reasonable to consider the AF-RVB wave function. However,
the mean field studies prefer RVB{\cite{zhang,gan}}. So we
included RVB in our work that it might get a comparison with the
mean field studies.  It is found that the RVB state has a lower
variational energy for smaller $U$ and $t_{mix}$ which is the
hopping amplitude changing the local double occupation while the
AF-RVB state is favorite for larger $U$ and $t_{mix}$. For both
wave functions, we check their phase structures. Both wave
functions have a crossover from metal to insulator states as
$U/t_{mix}$ tends to infinity. On the other hand, there is an
ambiguity to find an optimal variational paring parameter $\Delta$
describing the RVB feature of the states. In a wide range from
$\Delta=0$ to 1, the variational energies are almost degenerate
for the lattice size in our calculation. This leads to a
difficulty to identify if the metal state is either Fermi liquid
or $d$-wave superconducting.

To understand the physics in the crossover regime more clearly, we
study an effective model which includes the contribution up to the
second order of $t_{mix}/U$.  Moreover, the experience in the t-J
model taught us, if there is a spin exchange term in the
Hamiltonian, the pairing variational parameter is much easier to
be optimized \cite{odd}. For a large U, the spin exchange may be
explicitly shown by considering the virtual hopping of electron
between two single occupied sites. A hopping term changing the
double-occupancy may be taken as a perturbation as that in
deducing the t-J model from the single band Hubbard model. In the
perturbative deduction from the Hubbard model to the t-J model, we
see that the J-term is actually from the second order perturbation
in $t_{mix}/U$ by taking the off-diagonal term for the
double-occupied number in the kinetic term as the perturbation
Hamiltonian. Namely, if the t-term is decomposed into $T_{\rm
diag}+T_{\rm off-diag}$, the perturbation process transfers
$T_{\rm off-diag}$ to J-term, a virtual hopping process, and only
$T_{\rm diag}$ serves as the real hopping \cite{hir}. It can also
clearly be seen from the canonical transformation deduction of the
t-J model\cite{chao}. In the present case, we can still have a J
term as in the t-J model while the U term is kept due to a
non-zero double-occupancy. In the t-J model, due to the no-double
occupancy constraint, the kinetic term is a hopping between a
single-occupied site to an empty site. For the present model,
besides this hopping term, a hopping between double- and
single-occupied sites and a pair hopping between the
double-occupied and empty sites would be included. Thus, we can
derive an effective theory at large but finite U, which captures
both the charge and exchange correlations of the system. We can
have kinetic, J- and U- terms. However, it should not confuse with
the t-J-U model mentioned above. In t-J-U model, the J is set as a
free parameter that means J is independent of t and
U{\cite{daul}}. But in our case, J comes from the expansion of
$t_{mix}/U$. In this effective theory, there is no a hopping term
which changes the local double occupation. It is an extension of
the t-J model with doubly-occupied sites.

To work out our model, we use the canonical transformation. We
find that, to the second order of $t_{mix}/U$, the effective
Hamiltonian can be written as the sum over the Hamiltonians acting
on a subspace of the Hilbert space with a fixed double occupied
number $D$. This fixed $D$ Hamiltonian including three hopping
terms which serve the electron hopping from single to empty sites
($t_h$ term), double to single sites ($t_d$ term), and the paring
hopping ($t_p$ term); the U term and J-term as well as various
nearest neighbor interactions.

We also study the RVB and AF-RVB variational wave functions for
this effective model by the variational Monte Carlo method. The
RVB state is in smaller $J$ and $U$ regime while AF-RVB state is
favored in larger $J$ and $U$, consistent with the Hubbard model.
For both wave functions, a first order metal-insulator transition
may be found \cite{note1}. Finally, we can plot a phase diagram in
$J$-$U$ plane. The regime $J=4(t_{mix})^2/U$ with $t_{mix}/U\ll 1$
should describe the physics of the crossover regime in the Hubbard
model. We see that if neglecting the high order term, this
crossover corresponds to a metal-insulator phase transition.

This paper was organized as follows: In Sec. II, the detailed
deduction of the effective model is provided by canonical
transformations. In Sec. III, the VMC results for the original
Hubbard model and the effective model are presented. In Sec. IV,
we give some discussions and conclusions. The mean field theory is
arranged in the appendix for giving some feeling to relate our
effective model to Laughlin's gossamer superconducting model.

\section{Hubbard model and its large but finite effective model}

\subsection{One-band Hubbard Model}

We start from the Hubbard model on a two-dimensional square
lattice where the hopping energy may be dependent on the
occupation of sites involved \cite{AAA}. Including the on-site
Coulomb interaction, this Hubbard model reads
\begin{eqnarray}
H=T+V=T+U\sum_{i=1}^L\nu_i,
\end{eqnarray}
where $L$ is the number of the site;
$\nu_i=n_{i\uparrow}n_{i\downarrow}$,
$n_{i\sigma}=c^\dagger_{i\sigma}c_{i\sigma}$ with $c_{i\sigma}$ a
spin-$\sigma$ electron annihilation operator at site $i$ and the
kinetic term is given by
\begin{eqnarray}
&&T=T_h+T_d+T_{\rm mix},\\
&&T_h=-\sum_{\langle
ij\rangle\sigma}t^h_{ij}(1-n_{i\bar\sigma})c^\dagger_{i\sigma}
c_{j\sigma}(1-n_{j\bar\sigma}),\nonumber\\
&&T_d=-\sum_{\langle
ij\rangle\sigma}t^d_{ij}n_{i\bar\sigma}c^\dagger_{i\sigma}
c_{j\sigma}n_{j\bar\sigma},\nonumber\\
&&T_{\rm mix}=T_++T_-\nonumber\\
&&~~~~~~=-\sum_{\langle
ij\rangle\sigma}t^{mix}_{ij}n_{i\bar\sigma}c^\dagger_{i\sigma}
c_{j\sigma}(1-n_{j\bar\sigma})\nonumber\\
&&~~~~~~~~~-\sum_{\langle
ij\rangle\sigma}t^{mix}_{ij}(1-n_{i\bar\sigma})c^\dagger_{i\sigma}
c_{j\sigma}n_{j\bar\sigma}.\nonumber
\end{eqnarray}
Here $T_+(T_-)$ creates (destroys) a double-occupied site. We
assume $t^h_{ij}=t^d_{ij}=t$ and $t^{mix}_{ij}=t^m$ for the
nearest neighbor sites and vanish otherwise.

\subsection{Effective Model}

In large but finite $U$ ($U\gg t_{mix}$), we can treat the
$T_{mix}$ term as perturbation, which leads to the t-J model in
infinite $U$ limit. An easily pellucid way to arrive at the
effective model is via a canonical transformation. In order to
define the canonical transformation, we explain our notations. The
partial Gutzwiller projection operator
\begin{eqnarray}
\Pi(g)=\prod_i(1-(1-g)\nu_i) =\sum_{D=0}^{N/2}g^DP_D=g^{\hat D} ,
\end{eqnarray}
where $0\leq g\leq 1$ is the Gutzwiller parameters; $N$ is the
electron number \cite{note}, $\hat D=\sum_i\nu_i$ and
$$P_D=\sum_{\{i_1,...,i_D\}}[\nu_{i_1}...\nu_{i_D}\prod_j'(1-\nu_j)]
$$ is a projection operator which projects a state into the
subspace with a fixed double-occupation number $D$. $P_0=\Pi(0)$
is the full Gutzwiller projection operator and $\Pi(1)=1$. For
convenience, we denote $$ P_D(g)=g^DP_D, P_{\eta_i}(g)=\sum_{D\geq
i}P_D(g).$$

The first goal of this work is to construct an effective
Halmitonian $H_{\rm eff}$ and after the partial Gutzwiller
projection, the projected effective Hamiltonian is given by
\begin{eqnarray}
\Pi(g)H_{\rm eff}\Pi(g)&=&\sum_D P_D(g)H_{\rm
eff}P_D(g)\nonumber\\
&=&\sum_D g^{2D}P_DH_{\rm eff}P_D, \label{ham}
\end{eqnarray}
i.e., all the off-diagonal part $P_{D'}(g)H_{\rm eff}P_D(g)=0$ for
$D'\ne D$. We shall prove that the result effective Hamiltonian in
which all terms keep $D$-invariance is given by
\begin{eqnarray}
H_{\rm eff}=T_h+T_d+T_p+{\cal J}+V, \label{Heff}
\end{eqnarray}
where $T_p$ is a pair hopping kinetic energy and $\cal J$ is the
spin exchange as well as various nearest neighbor interactions,
namely,
\begin{eqnarray}
T_p&=&-\sum_{\langle
ij\rangle,\sigma}t_pc^\dagger_{i\sigma}c_{j\sigma}
c^\dagger_{i\bar\sigma}c_{j\bar\sigma},\nonumber\\
 {\cal
J}&=&\sum_{\langle ij\rangle} J_{ij}({\bf S_i}\cdot{\bf
S_j}-\frac{1}{4}n_in_j+\frac{1}{2}n_{i\uparrow}n_{i\downarrow}n_j\nonumber\\
&+&\frac{1}{2}n_in_{j\uparrow}n_{j\downarrow}
-n_{i\uparrow}n_{i\downarrow}n_{j\uparrow}n_{j\downarrow} ).
\end{eqnarray}
For $t_{mix}/U\ll 1$, $J_{ij}=J\approx 4t_{mix}^2/U,~~t_p=J$.

The canonical transformation for the Hubbard model to the t-J
model has been a standard technique \cite{chao}. A detailed review
for the canonical transformation can be found in Ref.\cite{bara}.
Our derivation is a generalization of the $D=0$ case. Notice that
$P_DT_{\rm mix} P_{D'}=\delta_{D',D\pm 1 }P_DT_{\rm mix}P_{D\pm
1}$ and (\ref{Heff}) remains $D$ invariant, as well as $
\Pi(g)\Pi(g')=\Pi(gg'),~~~P_{\eta_D}(g)P_{\eta_D}(g')
=P_{\eta_D}(gg')$. Keeping these in mind, we do a partial
projection $\Pi(x)H\Pi(x)$ with $x=g^{2/N}$. For large $N$, $x$ is
very close to 1. A straightforward calculation leads to a
rewriting of $\Pi(x)H\Pi(x)$
\begin{eqnarray}
&&\Pi(x)H\Pi(x)=H_0(x)+H_\eta^{(1)}(x),\nonumber\\
&&H_0(x)=H_{diag}(x)+\sum_{D=2}H_\eta^{(D)}(x),
\end{eqnarray}
where
\begin{eqnarray}
&&H_{diag}(x)=\sum_{D=0}P_D(x)HP_D(x),\nonumber\\
&&H^{(D)}_\eta(x)=P_{D-1}(x)TP_D(x)+P_D(x)TP_{D-1}(x).
\end{eqnarray}
The purpose of the canonical transformation is to acquire an
effective Hamiltonian $H_{\rm eff}^{(1)}$ such that $P_0H_{\rm
eff}^{(1)}P_D=P_DH_{\rm eff}^{(1)}P_0=0$ for $D\ne 0$ to the
second order of $t/U$. This $H_{\rm eff}^{(1)}$ is defined by
\begin{eqnarray}
H_{\rm eff}^{(1)}=e^{i S^{(1)}}\Pi(x)H\Pi(x)e^{-i S^{(1)}}.
\end{eqnarray}
As well-known\cite{chao,bara}, $S^{(1)}$ is determined by the
self-consistent condition $$iH_\eta^{(1)}(x)+[H_0(x),S^{(1)}]=0$$
and thus the effective Hamiltonian reads
\begin{eqnarray}
H_{\rm eff}^{(1)}&=&H_0(x)+\frac{i}2[S^{(1)},H_\eta^{(1)}(x)]
\nonumber\\
&-&\frac{1}3[S^{(1)},[S^{(1)},H_\eta^{(1)}(x)]]+....\label{h1}
\end{eqnarray}
Solving the self-consistent condition , $H_{\rm eff}^{(1)}$ in a
large U is given by \cite{hir,bara}
\begin{eqnarray} &&P_0H_{\rm eff}^{(1)}P_0\approx
P_0HP_0-\frac{1}{U}P_0HP_{\eta_1}HP_0,\label{p0}\\
&&P_{\eta_1}(x)H_{\rm eff}^{(1)}P_{\eta_1}(x)\approx
P_{\eta_1}(x^2)HP_{\eta_1}(x^2)\nonumber\\
&&~~~~~~~~~~~~~~~~~~~~~~~+\frac{1}{U}P_{\eta_1}(x^2)HP_0HP_{\eta_1}(x^2)
.\label{pe}
\end{eqnarray}
The approximation '$\approx$' in (\ref{p0}) and (\ref{pe}) means
the exactness is up to the second order of $t/U$. Namely, the
third term in (\ref{h1}) has been neglected. In fact, the
off-diagonal part $P_0H^{(1)}_{\rm eff}P_{\eta_1}$ vanishes also
only up to the second order:
\begin{eqnarray} P_0H^{(1)}_{\rm
eff}P_{\eta_1}&=& P_0HP_{\eta_1}HP_0HP_{\eta_1}\nonumber\\
&\approx& \frac{1}{U^2}T_-T_+T_-\sim O(tJ) \label{offd}
\end{eqnarray}
 is of the third order. The second
terms of (\ref{p0}) and (\ref{pe})  may be calculated and given by
\begin{eqnarray}
&&-\frac{1}{U}P_0HP_{\eta_1}HP_0=- \frac{1}{U}P_0T_-T_+P_0 \approx
P_0{\cal J}P_0,\nonumber\\
&&\frac{1}{U}P_{\eta_1}HP_0HP_{\eta_1}=\frac{1}UP_{\eta_1}T_+T_-
P_{\eta_1}\approx P_{\eta_1}T_pP_{\eta_1}.
\end{eqnarray}
Thus, up to the second order, we have
\begin{eqnarray}
&&P_0H_{\rm eff}^{(1)}P_0\approx P_0H_{\rm eff}P_0,\label{p02}\\
&&P_{\eta_1}(x)H_{\rm eff}^{(1)}P_{\eta_1}(x)\approx
P_{\eta_1}(x^2)(H+T_p)P_{\eta_1}(x^2)\label{pe2}.
\end{eqnarray}
where the approximation '$\approx$', besides up to the second
order, also means the three and more sites processes are
neglected.

 If the non-double occupied
constraint is imposed, (\ref{pe2}) vanishes because it is related
to the double occupation. Eq.(\ref{p02}) gives rise to the common
t-J model. However, if the double occupation is allowed, we have
to deal with (\ref{pe2}). In fact, one can repeats the canonical
transformation to (\ref{pe2}). We would like to require an
effective Hamiltonian $H_{\rm eff}^{(2)}$ whose off-diagonal part
$P_1H_{\rm eff}^{(2)}P_D=P_DH_{\rm eff}^{(2)}P_1=O(tJ)$ for $D>1$.
For this purpose, one writes
\begin{eqnarray}
\Pi(x)H^{(1)}_{\rm eff}\Pi(x)=P_0H^{(1)}_{\rm eff}P_0+\tilde
H_0(x^2)+H^{(2)}_\eta(x^2),
\end{eqnarray}
where
$$\tilde H_0(x^2)=P_1(x^2)(H+T_p)P_1(x^2)
+P_{\eta_2}(x^2)(H+T_p)P_{\eta_2}(x^2).$$ We do a canonical
transformation and define
\begin{eqnarray}
H_{\rm eff}^{(2)}=e^{iS^{(2)}}\Pi(x)H^{(1)}_{\rm
eff}\Pi(x)e^{-iS^{(2)}},
\end{eqnarray}
where $S^{(2)}$ is required to satisfy $P_0S^{(2)}=S^{(2)}P_0=0$
such that $P_0\Pi(x)H^{(1)}_{\rm eff}\Pi(x)P_0$ is invariant under
the transformation and it is self-consistently determined by
$$iH^{(2)}_\eta(x^2)+[\tilde H_0(x^2),S^{(2)}]=0.$$ Hence, similar
to (\ref{h1}), one has
\begin{eqnarray}
 H_{\rm eff}^{(2)}=P_0H_{\rm
eff}^{(1)}P_0+ \tilde
H_0(x^2)+\frac{i}2[S^{(2)},H_\eta^{(2)}(x^2)]+....
\end{eqnarray}
Projecting $H_{\rm eff}^{(2)}\to\Pi(x)H_{\rm eff}^{(2)}\Pi(x)$ and
repeating the similar procedure to deduce (\ref{p02}) and
(\ref{pe2}), one arrives at
\begin{eqnarray}
&&P_0H_{\rm eff}^{(2)}P_0\approx P_0H_{\rm eff}P_0,\nonumber\\
&&P_1(x)H_{\rm eff}^{(2)}P_1(x)\approx x^2P_1H_{\rm eff}P_1,\nonumber\\
&&P_{\eta_2}(x)H_{\rm eff}^{(2)}P_{\eta_2}(x)\approx
P_{\eta_2}(x^3)(H+T_p)P_{\eta_2}(x^3)\nonumber ,
\end{eqnarray}
for a large U, where the three site processes have been ignored.

Repeating this procedure, we finally have
\begin{eqnarray}
\Pi(x)H^{(\frac{N}2)}_{\rm eff}\Pi(x)&\approx&\sum_{D=0}
g^{2D}P_DH_{\rm eff}P_D\nonumber\\ &=&\Pi(g)H_{\rm eff}\Pi(g).
\label{hm}
\end{eqnarray}
The last equality is because $H_{\rm eff}$ is $D$-invariant. The
Gutzwiller parameter is $g$ but not $x$ because we are doing the
partial projection in each time canonical transformation. Thus, we
end the proof of (\ref{ham}) and (\ref{Heff}). Moreover, we see
that, in a partial Gutzwiller projection, the variational ground
state energy is given by a polynomial of the Guztwiller parameter
$g$ in power of $2D$. The coefficient of $g^{2D}$-term is the
ground state energy of the system with a fixed $D$. Using $g$ as a
variational parameter may be convenient for the numerical
simulations. In the original Hubbard model, the change of the
double occupied number is allowed. We see here that the allowance
of this change in a large U is very small. After neglecting the
three and  more sites processes, the probability of the change of
$D$ is in the third order of $t/U$ as eq. (\ref{offd}) shown.
Considering the fixed $D$ processes  may be helpful to numerical
simulations.

\section{Variational Monte carlo results}

\subsection{Variational Wave Functions}

The variational wave functions we would like to study are
so-called the partially projected RVB state $|\psi_D\rangle=P_D|
BCS\rangle$ and the partially projected AF-RVB state
$|\psi_D\rangle=P_D|AF$-$BCS\rangle$. The BCS state is defined by
\begin{eqnarray}
| BCS\rangle=\prod_k(u_k+v_kc^\dagger_{k\uparrow}
c^\dagger_{-k\downarrow})|0\rangle,
\end{eqnarray}
where $u_k$ and $v_k$ follows the standard BCS form
\begin{eqnarray}
&&a(k)=\frac{v_k}{u_k}=\frac{\Delta_k}{\xi_k+E_k},\\
&&\xi_k=-2(\cos k_x+\cos
k_y)-\mu,~~E_k=\sqrt{\xi_k^2+\Delta_k^2},\nonumber
\end{eqnarray}
for the $d$-wave pairing parameter $\Delta_k=\Delta(\cos k_x-\cos
k_y)$. The AF-BCS coexisted state $|AF-BCS\rangle$ is defined by
\cite{og}
\begin{eqnarray}
&&|AF-BCS(\Delta_d,\Delta_{af},\mu)\rangle
=\prod_{k,s}(u^{(s)}_k+v_k^{(s)}
d^{(s)\dagger}_{k\uparrow}d^{(s)\dagger}_{-k\downarrow})|0\rangle
\nonumber\\
&&\propto\exp\biggl[\sum_{k,s}\frac{v^{(s)}_k}{u^{(s)}_k}
d^{(s)\dagger}_{k\uparrow}d^{(s)\dagger}_{-k\downarrow}\biggr]
|0\rangle,
\end{eqnarray}
where
\begin{eqnarray}
\tilde a^{(\pm)}_k=\frac{v^{(\pm)}_k}{u^{(\pm)}_k}
=\frac{\pm\Delta_d\gamma_k}{(\pm E_k-\mu)+\sqrt{(\pm E_k-\mu)^2+
(\Delta_d\gamma_k)^2}},
\end{eqnarray}
and $E_k=\sqrt{\epsilon_k^2+\delta_{af}^2}$, $\epsilon_k=-2(\cos
k_x+\cos k_y)$ and $\gamma_k=2(\cos k_x-\cos k_y)$; and
\begin{eqnarray}
&&d^{(+)\dagger}_{k\sigma}=\alpha_{k\sigma}c_{Ak\sigma}-\beta_{k\sigma}
c_{Bk\sigma},\nonumber\\
&&d^{(-)\dagger}_{k\sigma}=\beta_{k\sigma}c_{Ak\sigma}+\alpha_{k\sigma}
c_{Bk\sigma},
\end{eqnarray}
with
\begin{eqnarray}
\alpha_{k\sigma}&=&\sqrt{\frac{1}2\biggl(1-\frac{\sigma\Delta_{af}}
{E_k}\biggr)},\nonumber\\
\alpha_{k\sigma}&=&\sqrt{\frac{1}2\biggl(1+\frac{\sigma\Delta_{af}}
{E_k}\biggr)}.
\end{eqnarray}
$c_{Ak\sigma} (c_{Bk\sigma})$ is the electron operator on
sublattice A(B).

\subsection{Hubbard Model with Asymmetric Hopping}

We first make a variational calculation for the original Hubbard
model. The energy we want to minimize is given by
\begin{eqnarray}
E_H&=&Ud+\frac{\sum_D y^DN_D (T_{h,D}+T_{d,D})}{\sum_D y^DN_D}\label{EH}\\
&+&\frac{\sum_Dy^{D+1/2}\sqrt{N_DN_{D+1}}T^{mix}_{D,D+1}}{\sum_D
y^DN_D},\nonumber
\end{eqnarray}

%(T_{+,D+1,D}+T_{-,D,D+1})
where $y=g^2$, $N_D=\langle\psi_D|\psi_D \rangle$ for the
partially projected RVB state $|\psi_D\rangle=P_D| BCS\rangle$ or
the partially projected AF-RVB state
$|\psi_D\rangle=P_D|AF-BCS\rangle$.

The average double occupation number $d$ is given by
\begin{eqnarray}
d=\frac{\sum_{D=0} y^{D}N_D D/L}{\sum_{D=0} y^{D}N_D}.\label{d}
\end{eqnarray}
And
\begin{eqnarray}
&&T_{h(d),D}=\frac{\langle\psi_D|T_{h(d)}|\psi_D\rangle}{N_D},\nonumber\\
&&T^{mix}_{D,D+1}=\frac{\langle\psi_{D+1}|T_+|\psi_D\rangle}{\sqrt{N_DN_{D+1}}}
+\frac{\langle\psi_D|T_-|\psi_{D+1}\rangle}{\sqrt{N_DN_{D+1}}}.
\end{eqnarray}

Let $\{|\alpha_D\rangle\}$ be a set of the basis in the
configuration space with a fixed $D$. The normal factors $N_D$ is
given by
\begin{eqnarray}
&&N_D=\sum_{\alpha_D}~\langle\psi_D|\alpha_D\rangle\langle\alpha_D|\label{nd}
\psi_D\rangle\nonumber \nonumber\\
&&=\sum_{\alpha_D}
|\langle\alpha_D|\psi_D\rangle|^2=\sum_{\alpha_D}
|A_{\alpha_D}|^2,
\end{eqnarray}
where $A_{\alpha_D}$ is just the determinant of the configuration
$\alpha_D$. We are not able to calculate $N_D$ exactly. We use the
approximation by taking all probabilities $|A_{\alpha_D}|^2$ to be
the same \cite{vol,OKM}. Thus, at half-filling
\begin{eqnarray}
N_D^{RVB}=\frac{L!}{[(N/2-D)!]^2D!(L-N+D)!}\label{nd1},
\end{eqnarray}
for the RVB case. In the AF-RVB case, the lattice is divided into
two sublattices A and B respectively and
\begin{eqnarray}
&&N_D^{AF-RVB}=\sum_{N_{A\uparrow}N_{A\downarrow}N_{AD}N_{AE}}\nonumber\\
&& \frac{(L/2)!}
{N_{A\uparrow}!N_{A\downarrow}!N_{AD}!N_{AE}!}\frac{(L/2)!}
{N_{B\uparrow}!N_{B\downarrow}!N_{BD}!N_{BE}!}\label{nd2},
\end{eqnarray}
where the configurations
$(N_{A\uparrow}$,$N_{A\downarrow}$,$N_{AD}$,$N_{AE}$;$N_{B\uparrow}$,
$N_{B\downarrow}$,$N_{BD}$,$N_{BE})$ are corresponding to numbers
of spin-up, spin-down, double occupancy and empty sites for each
sublattice and subjected to the following constraints:
\begin{eqnarray}
N_{A\uparrow}+N_{A\downarrow}-N_{AD}+N_{AE}=\frac{N}{2}\nonumber\\
N_{B\uparrow}+N_{B\downarrow}-N_{BD}+N_{BE}=\frac{N}{2}\nonumber\\
N_{A\uparrow}+N_{B\uparrow}=N_{A\downarrow}+N_{B\downarrow}=\frac{N}{2}\nonumber\\
N_{AD}+N_{BD}=N_{AE}+N_{BE}=D
\end{eqnarray}
\begin{figure}[htb]
\begin{center}
\includegraphics[width=6cm]{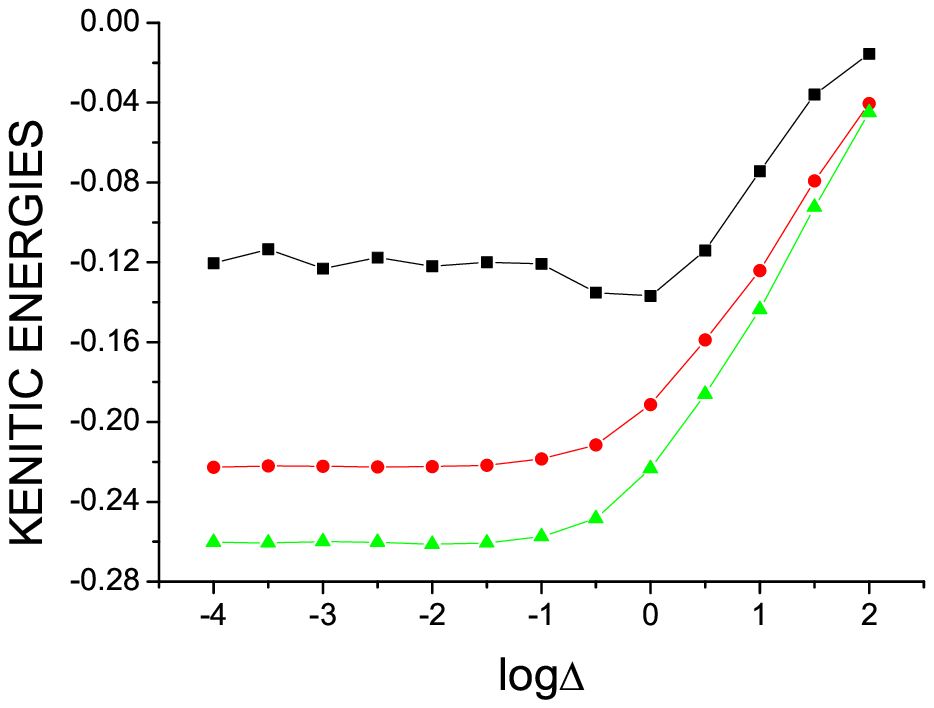}
\centerline{(a)}
\includegraphics[width=6cm]{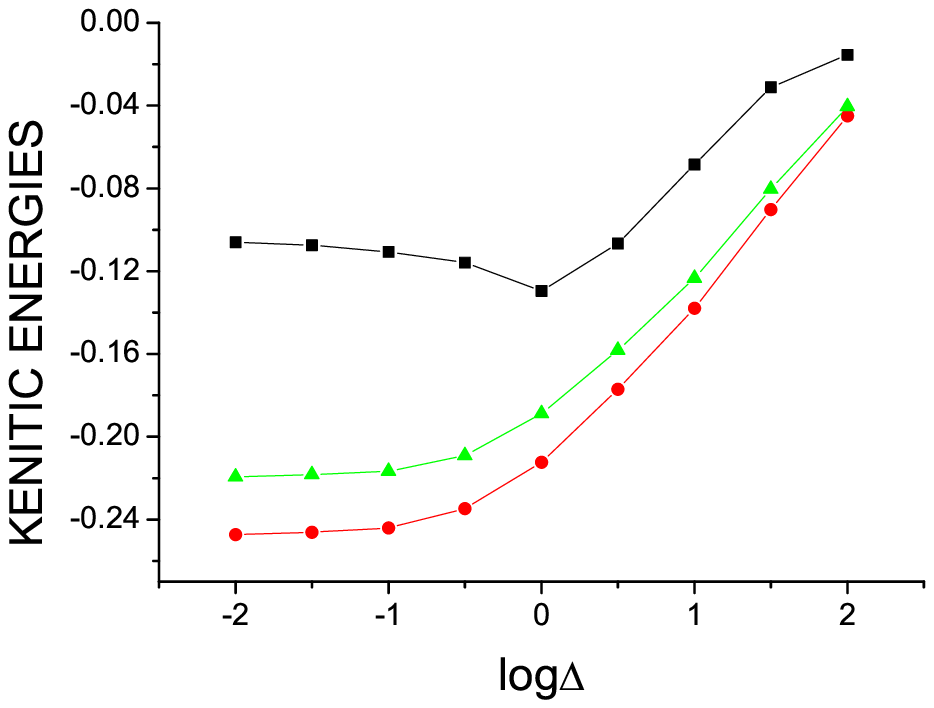}
\centerline{(b)}
\includegraphics[width=6cm]{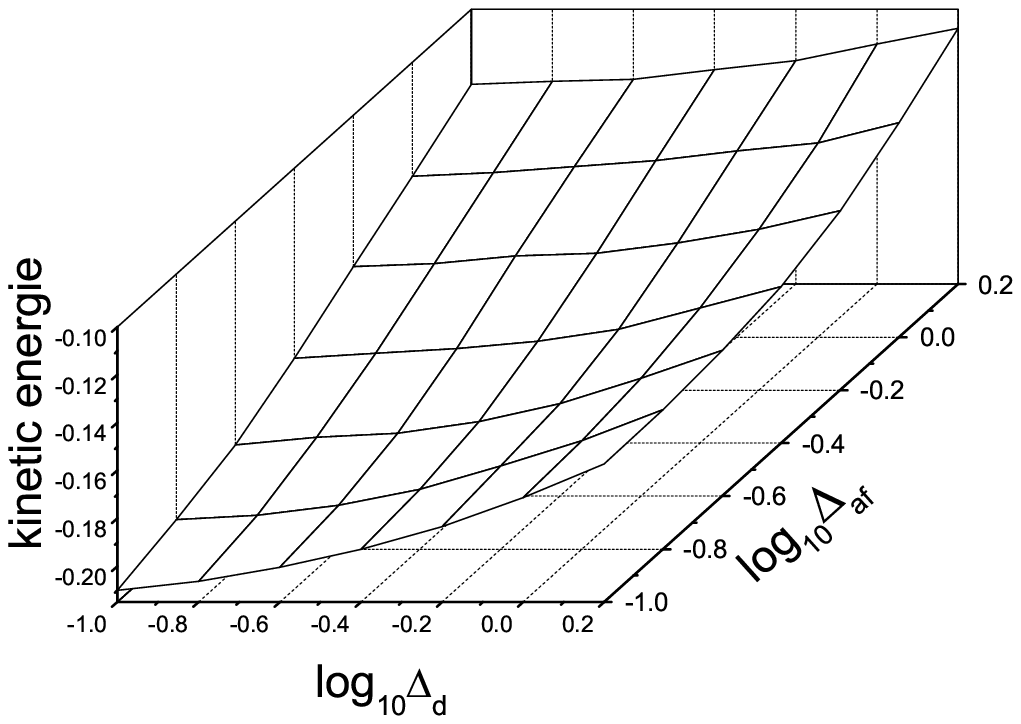}
\centerline{(c)}
\includegraphics[width=6cm]{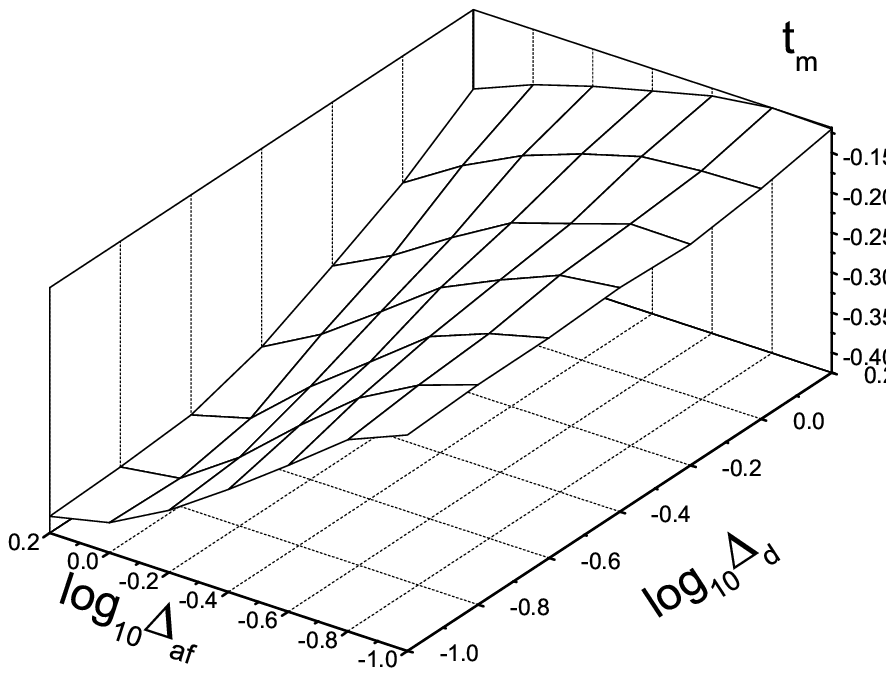}
\centerline{(d)}
\end{center}
\caption{ (a)(b)The kinetic energies $T_{h,D} +T_{d,D}$ and
$T_{mix, D,D+1}$ of RVB as functions of ${\rm log}_{10}\Delta$.
(a) $T_{-,5,6}+T_{+,6,5}$ (squares), $T_{h,5}+T_{d,5}$ (circles)
and $T_{h,6}+T_{d,6}$ (triangles) in a $10\times 10$ lattice. (b)
$T_{-,7,8}+T_{+,8,7}$ (triangles), $T_{h,7}+T_{d,7}$ (squares) and
$T_{h,8}+T_{d,8}$ (circles) in a $12\times 12$ lattice.(c)(d)The
kinetic energies $T_{h,D} +T_{d,D}$ and $T_{mix, D,D+1}$ of AF-RVB
as functions of ${\rm log}_{10}\Delta_d$ and ${\rm
log}_{10}\Delta_{af}$ in a $10\times 10$ lattice.(c)$T_{h,5}
+T_{d,5}$.(d) $T_{-,5,6}+T_{+,6,5}$. Note that for the different
trend between  $T_{h,D} +T_{d,D}$ and $T_{mix, D,D+1}$, we change
the view of $T_{mix, D,D+1}$.}\label{fig1}
\end{figure}

By using the variational Monte Carlo method \cite{vmc}, we
calculate the variational energy (\ref{EH}) by optimizing the
variational parameter $\Delta$.  The term $Ud$ is not dependent on
$\Delta$. For the projected RVB wave function, $T_{h,D},T_{d,D}$
and $T^{mix}_{D,D+1}$ for several $D$ are depicted in Fig.
\ref{fig1}. (The energy unit $t=1$ is used in all figures through
the paper.) The lattice sizes are $10\times 10$ and $12\times 12$,
respectively. We use periodic-antiperiodic boundary condition to
avoid the degeneracy in Brillouin zone. All data are calculated
with more than 10$^4$ Monte Carlo samples. Although there is a
minimum in $T^{mix}_{D,D+1}$ around ${\rm log}_{10}\Delta=0$, the
total kinetic energy is minimized after ${\rm
log}_{10}\Delta<-1.0$ because the minima of $T_{h(d),D}$ are in
after ${\rm log}_{10}\Delta<-1.0$. Unfortunately, we see that
there is very broad minimal flat in variational energy from
$\Delta=0$ to ${\rm log}_{10}\Delta\approx -1.0$. Thus, we can not
distinguish the metal state is either the Fermi liquid or
superconducting state. For the AF-RVB wave function, the trend of
$T_{h,D} +T_{d,D}$ and $T_{mix, D,D+1}$ is different. However, in
the total energy {\ref{EH}} $T_{h,D} +T_{d,D}$ dominate. So  the
situation is like the RVB case. The parameters we use are
$\log_{10}\Delta=\log_{10}\Delta_d=-0.6$ and the optimal
$\log_{10}\Delta_{af}\approx -0.6$ \cite{odd,ogata}.

\begin{figure}[htb]
\begin{center}
\includegraphics[width=6cm]{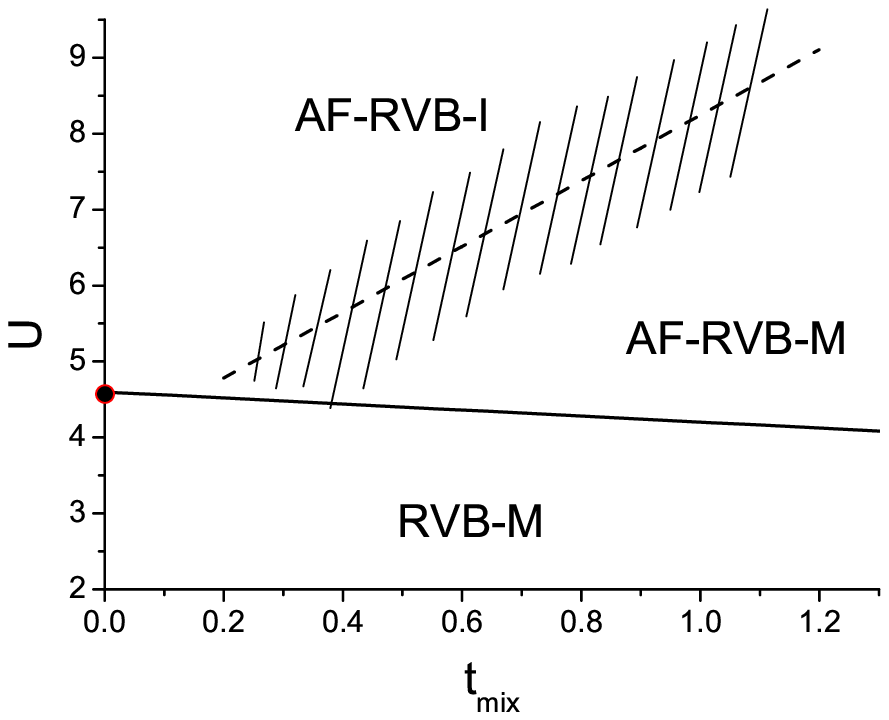}
\end{center}
\caption{The possible phase diagram of the Hubbard model. The
solid line divides the $t_{mix}$-$U$ plane into the RVB and AF-RVB
regions. The RVB region is in metal state (RVB-M). There is a
crossover from metal to insulator in AF-RVB region(AF-RVB-M to
AF-RVB-I). The spot on the U-axis ($t_{mix}=0$) is the
metal-insulator phase transition point $U_c(0)$. The shade area is
the crossover region and along the dashed curve,
$d_0=0.01$.}\label{fig2}
\end{figure}

Compare the variational energies of the two wave functions, we
find that the system is in the projected RVB state for small $U$
and $t_{mix}$ while it is in the projected AF-RVB coexisted state
for larger ones. Table{\ref{tab1}} shows the transition when
$T_{mix}=0.6$. The critical line is shown in Fig. \ref{fig2}.
\begin{table*}
\begin{tabular}{c|ccccccc|c}
  U            &0      & 1      & 2      & 3      & 4      & 5       & 6        &4.316\\
\hline
  $E_{RVB}$    & -0.835 & -0.603 & -0.405 & -0.242 & -0.115 & -0.0245 & -0.00202 & -0.8324 \\
  $E_{AF-RVB}$ & -0.795 & -0.564 & -0.373 & -0.221 & -0.109 & -0.0425 & -0.0205  &-0.8324\\
\end{tabular}
\caption{The transition of Hubbard model between RVB and AF-RVB at
t$_{mix}$=0.6. The critical U is 4.316.}\label{tab1}
\end{table*}

\begin{figure}%[htb]
\begin{center}
\includegraphics[width=6cm]{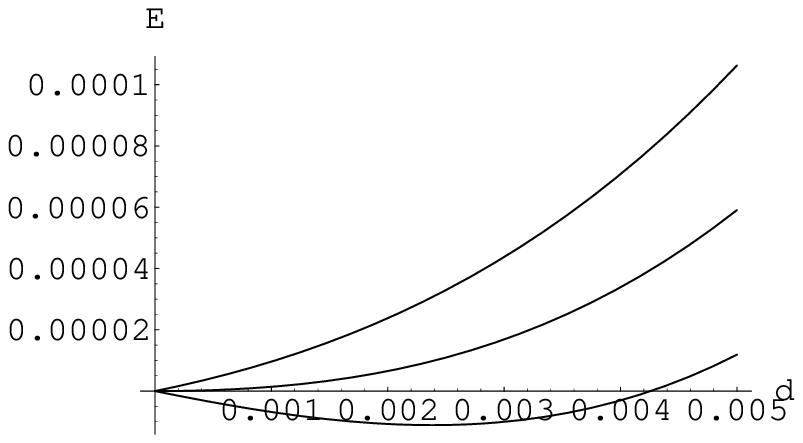}
\centerline{(a)}
\includegraphics[width=6cm]{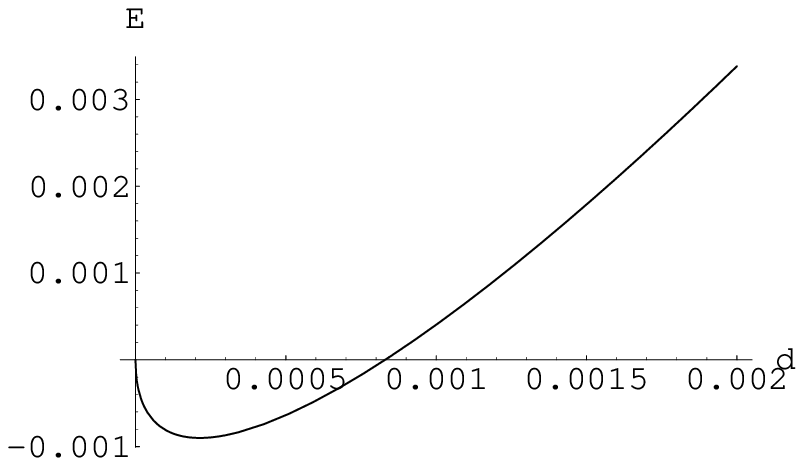}
\centerline{(b)}
\includegraphics[width=6cm]{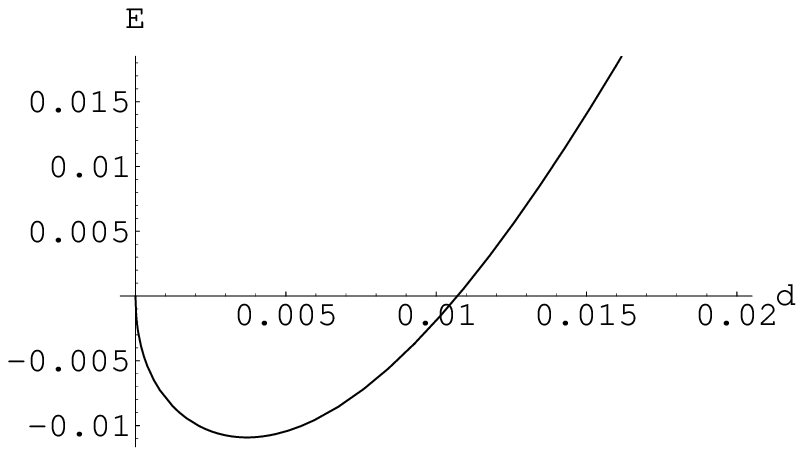}
\centerline{(c)}
\end{center}
\caption{ The $d$-$E$ curves of Hubbard model. (a) $t_{mix}=0$.
The different curves are corresponding to different $U$. Lower
curve has a smaller $U$.  The second phase transition happens at
$U_c(0)=4.5678$. (b) and (c) are $d$-$E$ curves for RVB state and
AF-RVB state at $t_{mix}=0.8$ and $U=10$,
respectively.}\label{fig3}
\end{figure}

To understand the phase diagram of the system, we shall calculate
the optimal average double occupied number $d$ for an appropriate
wave function (RVB or AF-RVB) for given $t_{mix}$ and $U$ in the
optimal parameters ($\Delta$ or $\Delta_d$ and $\Delta_{af}$).
Substituting (\ref{d}) into (\ref{EH}) and eliminating y, we get
the function E(d). Then, identifying the minimum of $E$ over $d$,
we get the optimal $d_0$ and $E_0$. If $d_0=0$, the system is in
insulating state while if $d_0>0$, the system is in metal state.
 There is a second order phase metal-insulator
transition in $t_{mix}=0$ as show by Fig. \ref{fig3}a. The
critical interaction $U_c(0)$ is spotted in Fig. \ref{fig2}.
However, when $t_{mix}>0$,
\begin{eqnarray}
\frac{\partial E(d)}{\partial d}|_{d\rightarrow 0}=U+(T_1-T_0)+
\frac{1}{C}\frac{1}{\sqrt{d}}T^{mix}_{0,1}
\end{eqnarray}
where C is a constant. For any finite $t_{mix}$, no matter how
large U is, it can be found there exists a $d_0>0$ so that
$\frac{\partial E(d_0)}{\partial d_0}=0$. As instances, in Fig.
\ref{fig3}(b)(c), we plot the $d$-$E$ curves for $t_{mix}=0.8$ and
$U=10$ for the RVB state (Fig. \ref{fig3}(b)) and the AF-RVB state
(Fig. \ref{fig3}(c)). The dashed curve in Fig. \ref{fig2} gives
the values of $(t_{mix},U)$ where the $d_0=0.01$. For a sufficient
small $d_0$, the system becomes a practical insulator and
therefore, there is a metal-insulator crossover as showed by the
shade area in Fig. \ref{fig2}. Due to small U, the RVB region is
in metal phase. The AF-RVB region is divided into two phases. For
a given $U$, the system is in the insulating phase when $t_{mix}$
is small enough while in metal state when $t_{mix}$ is large.

\subsection{Effective model}

We now begin to examine the effective model. The energy we want to
minimize is defined by
\begin{eqnarray}
E=Ud+\frac{\sum_{D}
y^{D}N_D(T_{h,D}+T_{d,D}+T_{p,D}+J_D)}{\sum_{D} y^{D}N_D}
\label{ed},
\end{eqnarray}
where d defined by (\ref{d}).

For fixed $D ,N_D$ defined by (\ref{nd}), (\ref{nd1}) and
 (\ref{nd2}). And $T_{h,D}, T_{d,D}, T_{p,D}, J_D$ is defined by
\begin{eqnarray}
&&T_{h(d,p),D}=\frac{\langle\psi_D|T_{h(d,p)}|\psi_D\rangle}{N_D},\nonumber\\
&&J_D=\frac{\langle\psi_D|J|\psi_D\rangle}{N_D}
\end{eqnarray}

Our strategy is that using the variational Monte Carlo method to
minimize $E_D$ for fixed $D$ and fixed electron number $N$ at the
half-filling by varying the variational parameter ${\rm
log}_{10}\Delta$. Then, draw the curves $E$ as the function of $d$
through eqs. (\ref{ed}) and (\ref{d}), to read out critical $U_c$
and $d_c$ from the shape of the curve for different model
parameters $U/t$ and $J/t$. At the moment, although we still use
$J=4t_{mix}^2/U$, we do not not restrict at $t_{mix}/U\ll 1$. The
comparison to the Hubbard model is only valid in the region
$t_{mix}/U\ll 1$.

\begin{figure}[htb]
\begin{center}
\includegraphics[width=6cm]{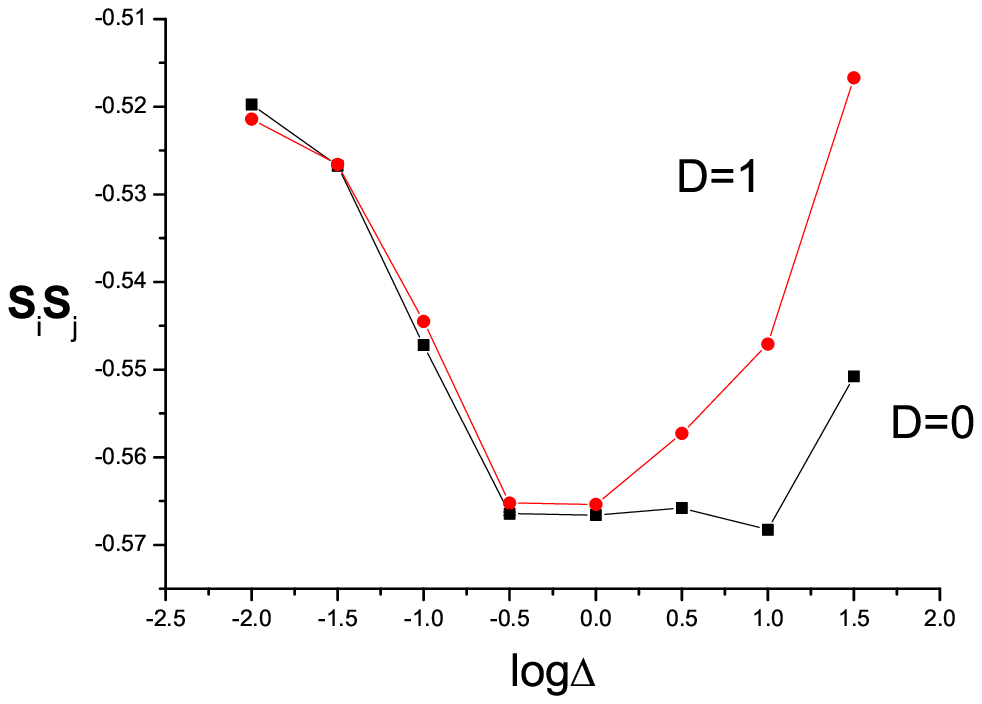}
\centerline{(a)}
\includegraphics[width=6cm]{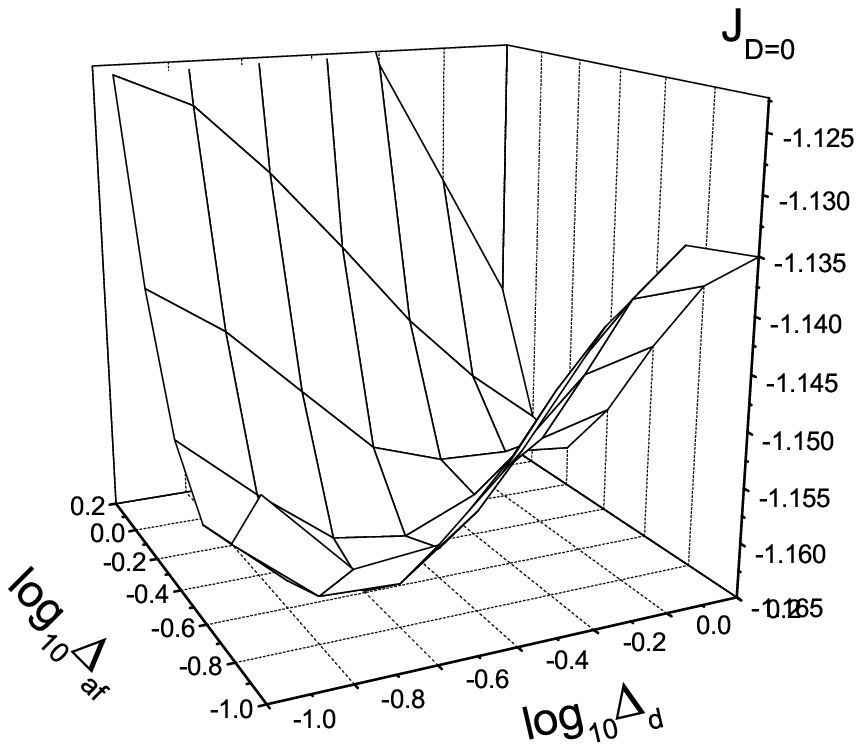}
\centerline{(b)}
\includegraphics[width=6cm]{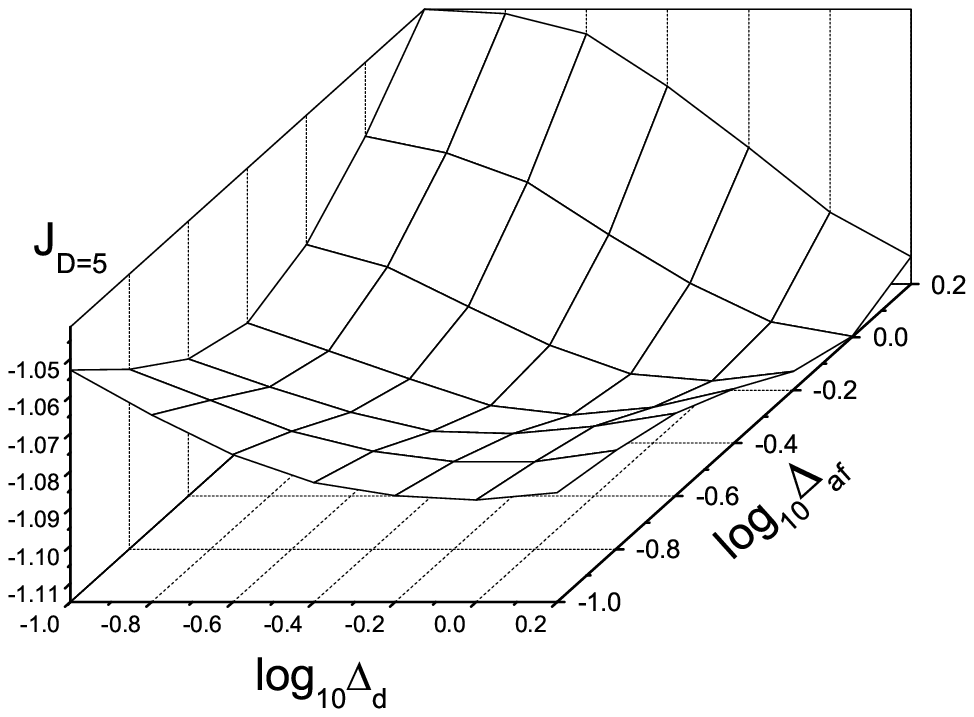}
\centerline{(c)}
\end{center}
\caption{The variational energy for fixed $D$ varying as the
variational parameter ${\rm log}_{10}\Delta$ for RVB(a), $ \rm
log_{10}\Delta_d$ and $\rm log_{10}\Delta_{af}$ for AF-RVB(b)(c).
}\label{fig4}
\end{figure}

Our variational Monte Carlo carries out on square lattices as in
the Hubbard model above, with sites $L$ from $10\times 10$ to
$12\times 12$. A periodic-antiperiodic boundary condition is used.
All data are calculated with more than 10$^4$ Monte Carlo samples.
In the half-filling, we set the chemical potential $\mu=0$. The
ground state energies $E_D$ are calculated. We show
$\vec{S_i}\cdot\vec{S_j}$ for D=0 and D=1 varying as ${\rm
log}_{10}\Delta$ in Fig.\ref{fig4}a for the RVB state. The
no-double occupant energy D=0 is the variational ground state
energy of the common $t$-$J$ model. Our result is well consistent
with the known results \cite{odd,vmc}. We calculate $E_D$ up to
the largest $D=L/2-1$, and find that all these energies are almost
degeneracy in wide range between $-0.5\leq {\rm log}_{10}\Delta
\leq 0.0$ . Using the Monte Carlo estimating energy $E_D$ on
$10\times 10$ lattice, we approximate $E$ in (\ref{ed}) by finite
sum for $D=49$ and ${\rm log}_{10}\Delta=-0.5$.  The error bars
for independent Monte Carlo initial configuration are in order of
$1\%$ and we do not show them.

 The energy of the AF-RVB wave function also can be calculated
by variational Monte Carlo method with optimizing both of the
model parameters $\Delta_{af}$ and $\Delta_d$.We show J$_D$ for
D=0 and D=5 in Fig.\ref{fig4}b and Fig.\ref{fig4}c. The results of
D=0 corresponding t-J model at half filling. Our results are
consistent with the known results.{\cite{og}} One can see that for
D=0, the energy minimum locates in a deep valley.

\begin{figure}[h]
\begin{center}
\includegraphics[width=6cm]{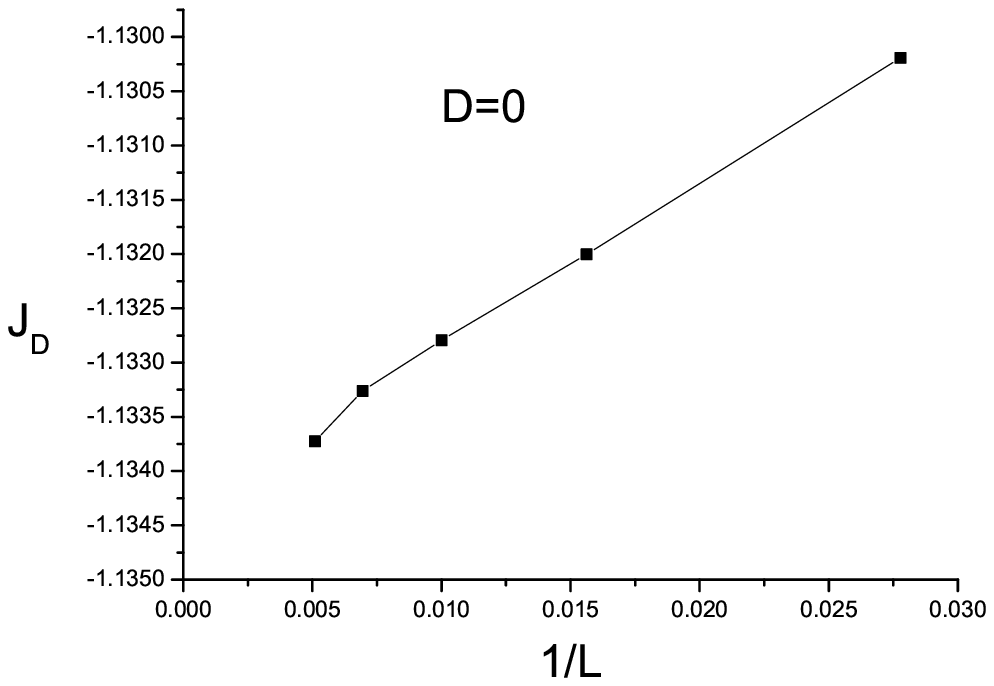}
\centerline{(a)}
\includegraphics[width=6cm]{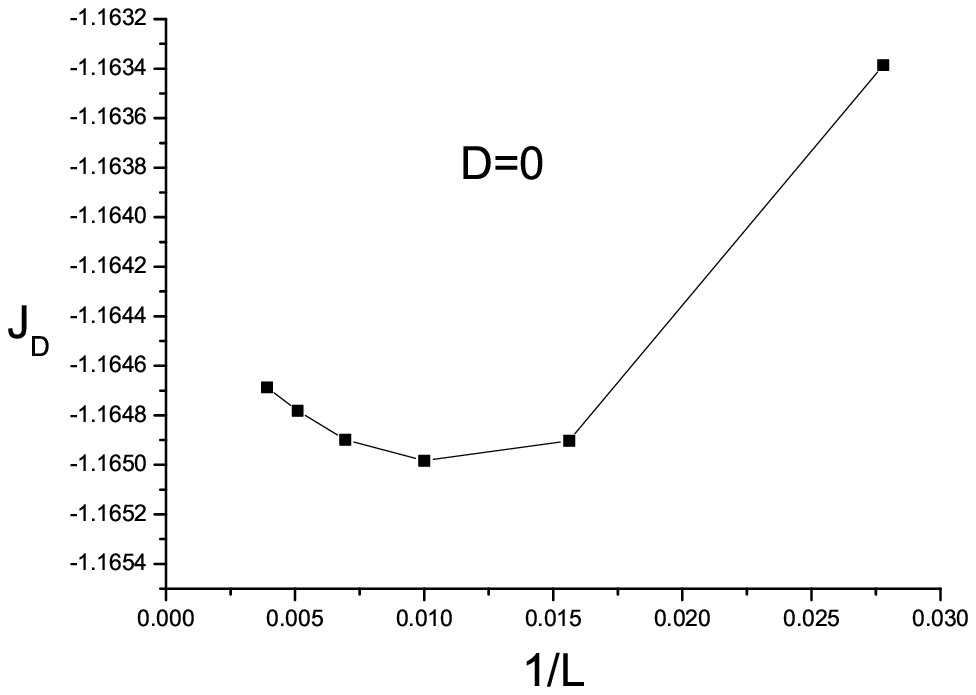}
\centerline{(b)}
\end{center}
\caption{Finite-size scaling of RVB at log$_{10}\Delta$=-0.5(a)
and AF-RVB at log$_{10}\Delta_d$=-0.6 and
log$_{10}\Delta_{af}$=-0.6(b).Where L is the lattices size.
}\label{fig5}
\end{figure}

We also analyze the two wave functions' finite-size scaling of D=0
which corresponds to Heisenberg model. The results are show in
Fig.\ref{fig5}. All the data but the 16$\times$16 of AF-RVB, which
is only one datum since it is required very long time to get one
result, are average of 5 independent calculations. One can see
that for Heisenberg model the energies of AF-RVB are deeper than
those of RVB.

\begin{figure}[htb]
\begin{center}
\includegraphics[width=6cm]{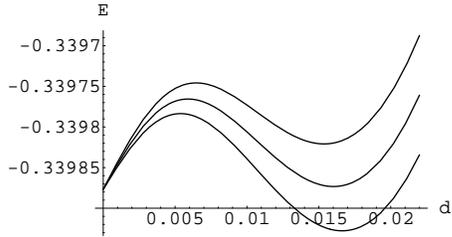}
\end{center}
 \caption{ The total energy $E$ averaged
in RVB variational wave function varying with the double occupant
concentrate $d$ for $J=0.3$
 and $U$=3.804, 3.8007, 3.7974 from the upmost to the lowest, respectively.
The critical $U_c=3.8007$ and $d_c=0.0135$ in this case.}
 \label{fig6}
\end{figure}

\begin{figure}
\begin{center}
\includegraphics[width=6cm]{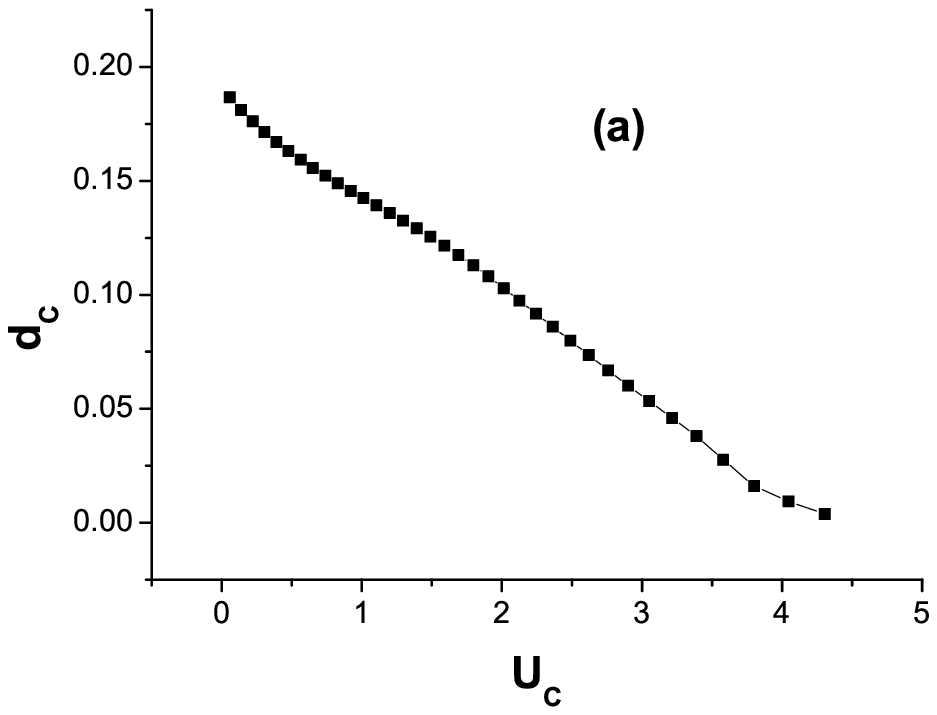}
\includegraphics[width=6cm]{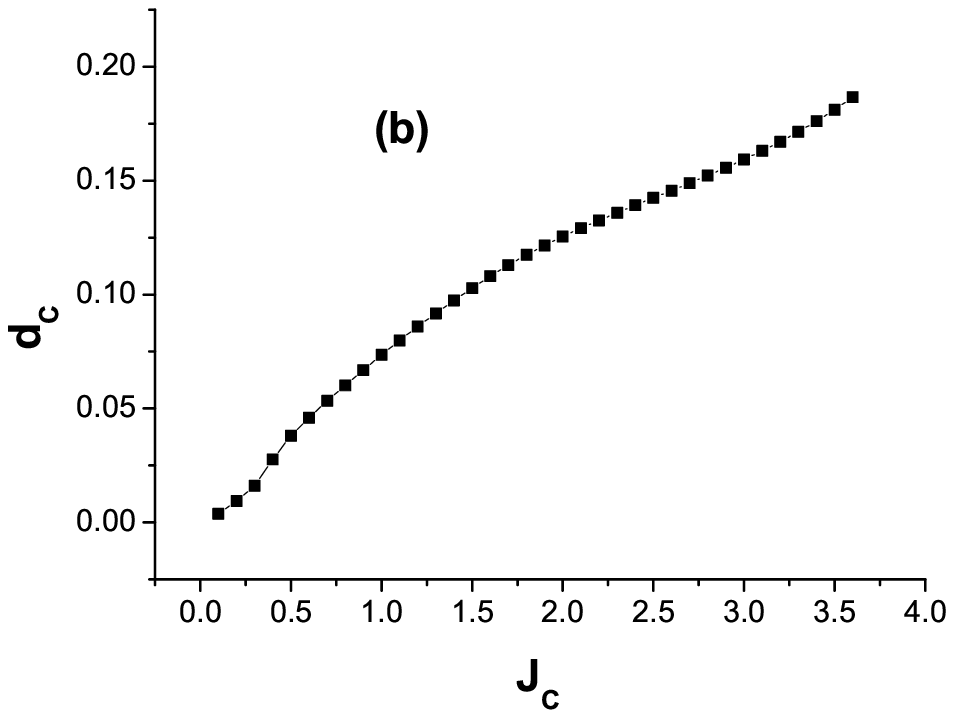}
\end{center}
\caption{The critical concentration $d_c$ for the RVB state (a)
$U_c$-$d_c$ curve; (b) $J_c$-$d_c$ curve.}\label{fig7}
\end{figure}

\begin{figure}[h]
\begin{center}
\includegraphics[width=6cm]{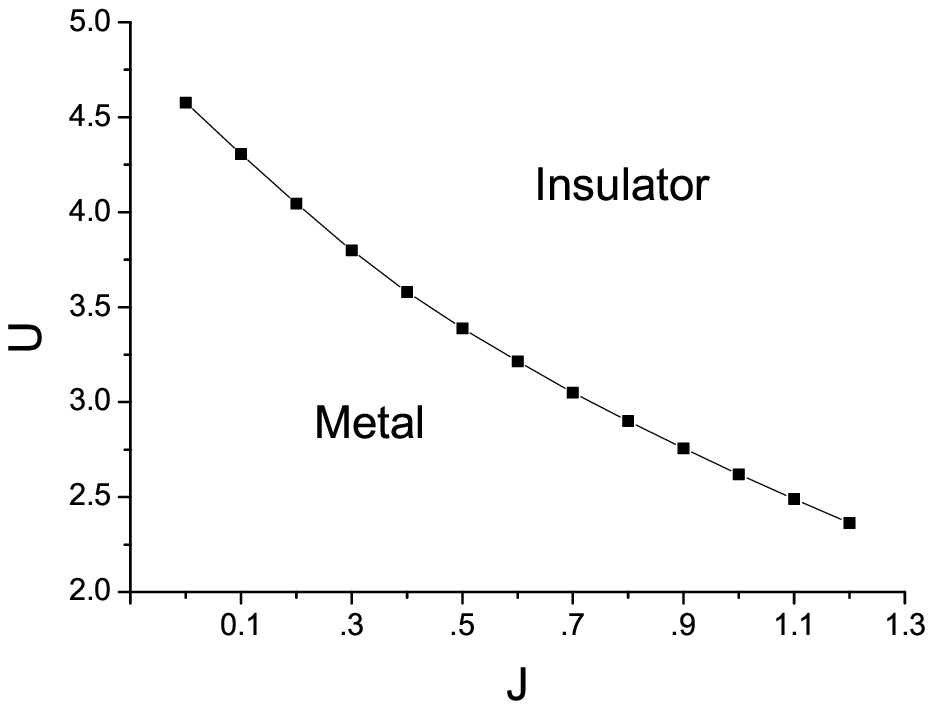}
\centerline{(a)}
\includegraphics[width=6cm]{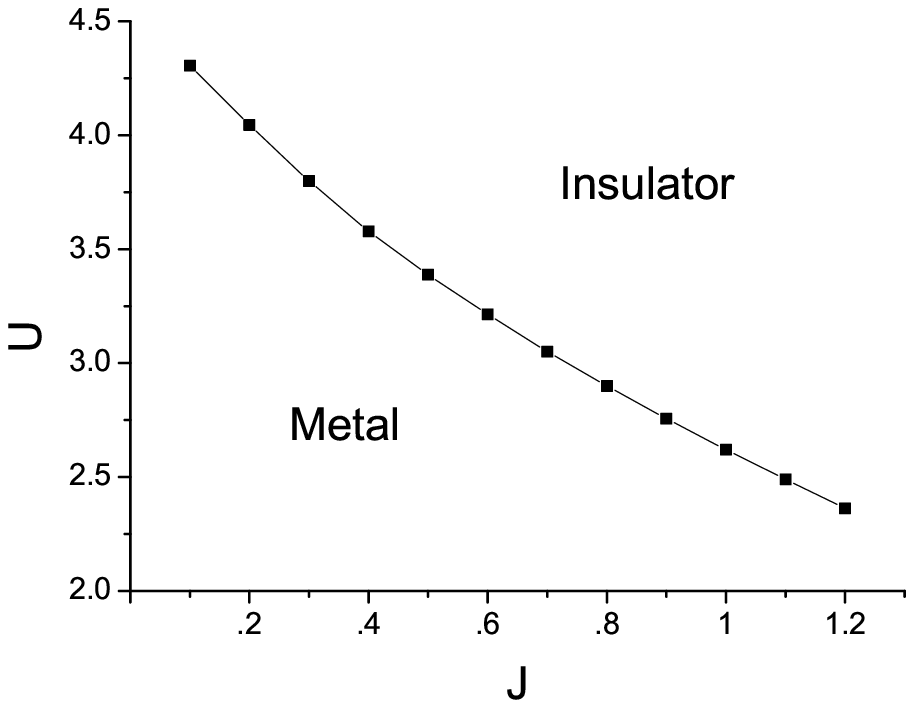}
\centerline{(b)}
\end{center}
\caption{ Phase diagrams for different variational wave functions.
(a) for the RVB wave function; (b) for the AF-RVB wave function
 }\label{fig8}
\end{figure}

\begin{figure}[h]
\begin{center}
\includegraphics[width=6cm]{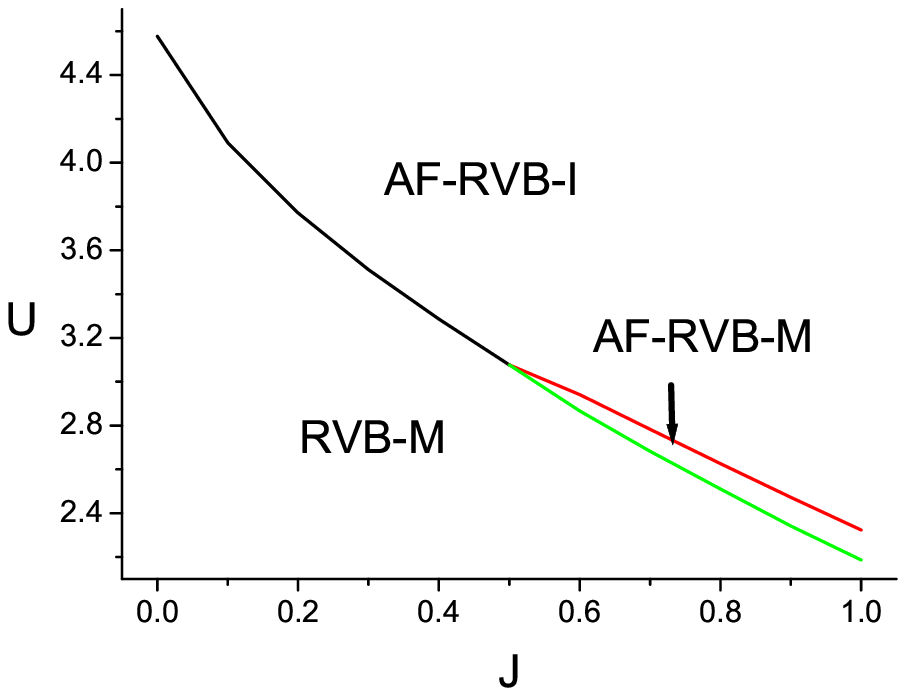}
\centerline{(a)}
\includegraphics[width=6cm]{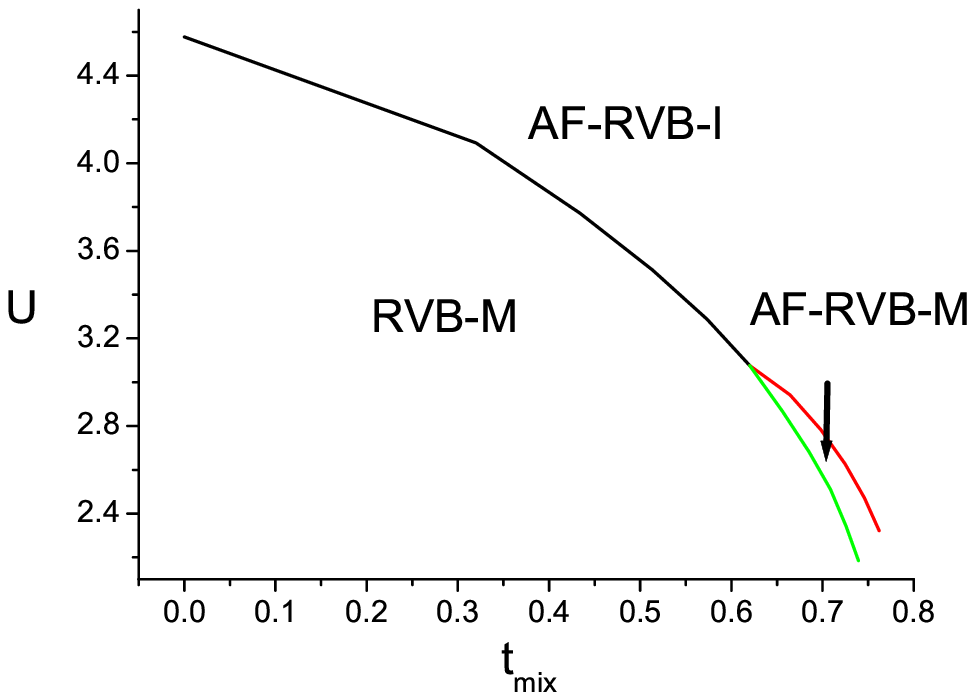}
\centerline{(b)}
\end{center}
\caption{ Phase diagrams of the effective model (a) the phase
diagrams of the system in the $J$-$U$ plane.(b) converting (a) to
$t_{mix}$-$U$. }\label{fig9}
\end{figure}

For a pair of fixed $J$ and $U$, we can compare the variational
energies corresponding to both wave functions(Tab.\ref{tab2}). In
this way, the $J$-$U$ plane can be divided into two regions: RVB
and AF-RVB, which is similar to the case in the last subsection
for the Hubbard model.
\begin{table*}
\begin{tabular}{c|cccccc|c}
  U            & 1      & 2      & 3      & 4      & 5       & 6        &3.512\\
\hline
  $E_{RVB}$    & -0.656 & -0.491 & -0.381 & -0.34    & -0.34    & -0.34    & -0.3494 \\
  $E_{AF-RVB}$ & -0.616 & -0.457 & -0.365 & -0.34943 & -0.34943 & -0.34943 &-0.3494\\
\end{tabular}
\caption{The transition of t-J-U model between RVB and AF-RVB at
J=0.3. The critical U is 3.512. The critical U of RVB M-I
transition is 3.8, and that of AF-RVB is 3.45.} \label{tab2}
\end{table*}

For a given type wave function, we look for the possible
metal-insulator transition. First, like the case in Hubbard model,
there is a second order phase transition when $J=0$. If $J>0$, due
to the vanishing of $T^{mix}$ term, there are first order phase
transitions in a given type wave function. Fig. \ref{fig6} gives
an example of the first order metal-insulator transition. In Fig.
\ref{fig8}, we show the relation between the critical $U_c$ or
$J_c$ and the critical double occupied concentration $d_c$ for the
RVB wave function. In this way, we can determine the critical
$J_c-U_c$ line in J-U plane. Figs. \ref{fig7}(a)(b) show the
critical $J_c-U_c$ lines for the RVB and AF-RVB wave functions.
They are quite similar. Combining these two phase diagrams
together with the region-dividing picture mentioned above, we
depict the comprehensive phase diagram (Fig. \ref{fig9}(a)). In
the RVB region, due to small J and U, the system is in metal
state. In the AF-RVB region, the system is basically in an
insulating phase. For $J \geq 0.5 $, there is a phase which may be
a AF-RVB metal state. Since the optimal variational parameters
$\Delta$  and $\Delta_d$ are not zero, the metal state may be a
superconducting state. Converting $J\to t_{mix}$ (see Fig.
\ref{fig9}(b)), we find that for small $t_{mix}$, the phase
diagram is consistent with the crossover picture in the Hubbard
model .

\section{ Discussions and conclusions}

We have investigated the Hubbard model with the hopping asymmetry
and deduced an effective theory for large but finite $U$. Based on
two types of the variational wave functions, the phase diagram of
both models are depicted by the variational Monte Carlo method.
For the Hubbard model, we found it is difficult to determine the
exact critical boundary of the phase transition of
metal-insulator. Moreover, the superconducting behavior in the
metal phase was not clear. The effective model is a finite but
large $U$ extension of the t-J model. This model captures both the
charge and exchange correlation. The phase diagram of this model
clearly shows a metal-insulator phase transition. Due to non-zero
optimal $\Delta$ and $\Delta_d$, the metal state may be
superconducting, which leads to the possibility of the gossamer
superconductivity in the framework of the hopping asymmetry
Hubbard model.

 The relation to the gossamer superconductivity can also be seen from
 the mean field state of our theory.
 The basic idea to go this mean field state has been explained in
 our previous preprint \cite{yu}.
 Here we present a renewed version of the
 mean field state. We only try to show our mean field theory may
 formally be equivalent to Laughlin's gossamer superconducting model.
 We do not intent to go more analysis such as the stability of
 our mean field state against other possible instabilities before
 we work out some more sophisticated issues. We put this formal
 identification into Appendix A.

\vspace{5mm}

{\centerline {\bf  ACKNOWLEDGEMENTS}}

\vspace{5mm}

The authors are grateful for the useful discussions to Jingyu Gan,
Jinbin Li, Zhaobin Su, Tao Xiang, Lu Yu and F. C. Zhang. One of
the authors (Y.W.) would like to thank Lei Zhang at C.I.T. for
some mathematical help. This work was supported in part by the NSF
of China. Part of the computation of this work was performed on
the HP-SC45 Sigma-X parallel computer of ITP and ICTS,CAS.

\appendix

\section{Mean field state}

 We outline the mean field state of our model in this
 appendix. Due to the paring hopping is of the order $J$, we
 neglected it in our mean field theory.
 Introducing two correlation functions $ \Delta_{ij}=\langle
c_{i\downarrow}c_{j\uparrow} - c_{i\uparrow}c_{j\downarrow}
\rangle_0,
 \chi_{ij}=\langle c^\dagger_{i\uparrow}c_{j\uparrow}
+c^\dagger_{i\downarrow}c_{j\downarrow} \rangle_0$, the $U(1)$
symmetry of $H_{\rm eff}$ is broken by a decomposition of the four
particle terms \cite{yu}. According to $\Delta_{ij}$ and
$\chi_{ij}$, the mean field Hamiltonian of (\ref{Heff}) is given
by
\begin{eqnarray}
H_{\rm MF}&=&-\sum_{\langle ij\rangle \sigma}
(t^h_{ij}+t^{(1)}_{ij}(n_{i\bar\sigma}+n_{j\bar\sigma})
+t_{ij}^{(2)}n_{i\bar\sigma}n_{j\bar\sigma})c^\dagger_{i\sigma}
c_{j\sigma} \nonumber \\
&+&\sum_{\langle ij\rangle\sigma}(J_{ij}+J_{ij}^{(1)}(n_{i\sigma}
+n_{j\bar\sigma})+J^{(2)}_{ij}n_{i\sigma}n_{j\bar\sigma})
\nonumber\\ &\times&(-1)^\sigma( \Delta^\dagger_{ij}
c_{i\sigma}c_{j\bar\sigma}+\Delta_{ij}
c^\dagger_{j\bar\sigma}c^\dagger_{i\sigma}) \nonumber\\
&+&U\sum_in_{i\uparrow}n_{i\downarrow}-\sum_{\langle
ij\rangle}2J_{ij}(1-A)n_{i\uparrow}n_{i\downarrow}\\
&-&\sum_{\langle
ij\rangle}J_{ij}(A|\Delta_{ij}|^2+\frac{1}{2}(1-B)|\chi_{ij}|^2)
(n_i+n_j)\nonumber\\ &+&\sum_{\langle
ij\rangle\sigma}J_{ij}(\frac{A}{2}
|\Delta_{ij}|^2n_{i\sigma}n_{j\bar\sigma}+\frac{1-B}{2}|\chi_{ij}|^2
n_{i\sigma} n_{j\sigma}),\nonumber
\end{eqnarray}
where the parameters are given by
\begin{eqnarray}
&&J^{(1)}_{ij}=\frac{A}{2}J_{ij},~~~J_{ij}^{(2)}=-\frac{B}2J_{ij},
\nonumber\\
&&t^{(1)}_{ij}=-t^h_{ij}-(1-A)\chi_{ji}J_{ij},\\
&&t^{(2)}_{ij}=t^h_{ij}+t_{ij}^d-(1-B)\chi_{ji}J_{ij}.\nonumber
\end{eqnarray}
$A$ and $B$ are the variantional parameters to be determined. On
the other hand, we write down Laughlin's gossamer superconducting
Hamiltonian
\begin{eqnarray}
H_G-\mu_RN=\sum_{\bf k}E_{\bf k}\tilde b^\dagger_{{\bf
k}\sigma}\tilde b_{{\bf k}\sigma}, \label{goh}
\end{eqnarray}
where $\mu_R$ is renormalized chemical potential,
$E_k=\sqrt{(\epsilon_k-\mu_R)^2+\Delta_k^2}$ and $\tilde b_{{\bf
k}\sigma}=\Pi(g)b_{{\bf k}\sigma}\Pi^{-1}(g)$ for $ b_{{\bf
k}\uparrow}=u_{\bf k} c_{{\bf k}\uparrow}+v_{\bf k}
c^\dagger_{-{\bf k}\downarrow}$ and $ b_{{\bf k}\downarrow}=u_{\bf
k} c_{{\bf k}\downarrow}-v_{\bf k} c^\dagger_{-{\bf k}\uparrow}$
annihilate the BCS state. Expressing explicitly (\ref{goh}) by the
electron operators \cite{yu,lau1}, we have
\begin{eqnarray}
&&H_G-\mu_R N=-\sum_{\langle ij\rangle
\sigma}[t^G_{ij}+t^{G(1)}_{ij}(n_{i\bar\sigma}+n_{j\bar\sigma})
\nonumber\\&&~~~+t^{G(2)}_{ij}n_{i\bar\sigma}n_{j\bar\sigma}]
+\sum_{\langle ij\rangle \sigma} J_{ij}
[1+\frac{1}2\alpha\beta(n_{i\bar\sigma}+n_{j\bar\sigma})\nonumber\\
&&~~~-\alpha\beta n_{i\bar\sigma}n_{j\bar\sigma}](-1)^\sigma
(\Delta_{ij}^\dagger c_{i\sigma}c_{j\bar\sigma}+\Delta_{ij}
c^\dagger_{j\bar\sigma}c^\dagger_{i\sigma})\nonumber\\
&&~~~+U_G\sum_in_{i\uparrow}n_{i\downarrow}-\mu_G N
\end{eqnarray}
where $\alpha=1-g$ and $\beta=(1-g)/g$ and
\begin{eqnarray}
t^G_{ij}~~&=&t_{ij}^h,\nonumber\\
t^{G(1)}_{ij}&=&-\sum_k\frac{E_{\bf k}}{M}(\alpha v_{\bf
k}^2+\beta u_{\bf k}^2)e^{i{\bf
k}\cdot({\bf r}_i-{\bf r}_j)}\nonumber\\
t_{ij}^{G(2)}&=&\sum_k\frac{E_{\bf k}}{M}(\alpha^2v_{\bf
k}^2-\beta^2u_{\bf k}^2)e^{i{\bf
k}\cdot({\bf r}_i-{\bf r}_j)}, \nonumber\\
J_{ij}\Delta_{ij}&=&\sum_{\bf k}\frac{E_{\bf k}}Mu_{\bf k}v_{\bf
k}e^{i{\bf k}\cdot({\bf r}_i-{\bf r}_j)},
\end{eqnarray}
and $ U_G=\frac{1}{M}\sum_{\bf k}E_k[(2\beta+\beta^2)u_k^2
+(2\alpha-\alpha^2)v^2_k],\mu_G=\frac{1}{M}\sum_{\bf
k}E_k[(2\alpha+1)v_k^2-u_k^2]$. If we identify the t-J-U model to
the gossamer superconducting model in the mean field level, one
requires
\begin{eqnarray}
&& A=\alpha\beta,~~~ B=2\alpha\beta, \nonumber\\ &&
t^{G(1)}=-t^h_{ij}-(1-\alpha\beta)\chi_{ji}J_{ij},\nonumber\\
&&t_{ij}^{G(2)}=t_{ij}^h+t^d_{ij}-(1-2\alpha\beta)\chi_{ji}J_{ij},
\end{eqnarray}
and $\mu_R+\mu_G=J(12A|\Delta_\tau|^2+8(1-B)|\chi_\tau|^2)+\mu,
U=U_G+8J(1-A)$.

Although we have made a formal equivalence between our mean field
state Hamiltonian to Laughlin gossamer superconducting
Hamiltonian, we note that the hopping parameters $t^{G(1,2)}$ have
run out of the practical range in the real materials. Thus, to
show the system described by the t-J-U model has a gossamer
superconducting phase described by Laughlin gossamer
superconducting Hamitonian, a renormalization group analysis is
required. We do not touch this aspect in this work. However, we
can believe there is such a superconducting phase in our theory if
$U< U_c$ because the superconducting paring parameter is
determined by the optimal exchange energy as in the common t-J
model. The renormalization of the hopping parameters is believed
to affect the normal dissipation process only.


\begin{references}
\bibitem{note} The summation over $D$ runs from 1 to $L$ in
general. However, for a fixed particle number $N$, the highest
number of the double occupied sites is $N/2$.
\end{references}
\end{document}